\documentclass[12pt]{article}
\usepackage{graphicx}
\usepackage{amssymb}
\usepackage{amsmath}
\usepackage{cite}
\newlength{\extraspace}
\newlength{\extraspaces}
\newcommand{\be}{\begin{equation}
\addtolength{\abovedisplayskip}{\extraspaces}
\addtolength{\belowdisplayskip}{\extraspaces}
\addtolength{\abovedisplayshortskip}{\extraspace}
\addtolength{\belowdisplayshortskip}{\extraspace}}
\newcommand{\ee}{\end{equation}}

\newcommand{\ba}{\begin{eqnarray}
\addtolength{\abovedisplayskip}{\extraspaces}
\addtolength{\belowdisplayskip}{\extraspaces}
\addtolength{\abovedisplayshortskip}{\extraspace}
\addtolength{\belowdisplayshortskip}{\extraspace}}
\newcommand{\ea}{\end{eqnarray}}

\newcommand{\newsection}[1]{
\vspace{12mm}
\pagebreak[3]
\addtocounter{section}{1}
\setcounter{equation}{0}
\setcounter{subsection}{0}
\noindent{\bf \thesection. #1}
\nopagebreak
\medskip
\nopagebreak}

\newcommand{\newsubsection}[1]{
\vspace{0.8cm}
\pagebreak[3]
\addtocounter{subsection}{1}
\noindent{\it \thesubsection. #1}
\nopagebreak
\vspace{2mm}
\nopagebreak}

\newcounter{saveeqn}

\newcommand{\dif}{\mathrm{d}}
\newcommand{\me}{\mathrm{e}}

\begin{document}
\addtolength{\baselineskip}{1.5mm}

\thispagestyle{empty}
\begin{flushright}

\end{flushright}
\vbox{}
\vspace{2cm}

\begin{center}
{\LARGE{Balanced electric-magnetic dihole \\[2mm]
in Kaluza--Klein theory
        }}\\[16mm]
{Yu Chen$^1$~~and~~Edward Teo$^{2,1}$}
\\[6mm]
$^1${\it Department of Physics,
National University of Singapore, %\\[1mm]
Singapore 119260}\\[5mm]
$^2${\it Centre for Gravitational Physics, College of Physical and Mathematical Sciences,\\[1mm]
The Australian National University, Canberra ACT 0200, Australia}\\[15mm]

\end{center}
\vspace{2cm}

\centerline{\bf Abstract}
\bigskip
\noindent
We present a four-dimensional double-black-hole (or dihole) solution in Kaluza--Klein theory, describing a superposition of an electrically charged and a magnetically charged black hole. This system can be balanced for appropriately chosen parameters, and the resulting space-time is completely regular on and outside the event horizons. This solution was constructed using the inverse-scattering method in five-dimensional vacuum gravity, in which it describes a rotating black ring surrounding a static black hole on a Taub-NUT background space. Various properties of this solution are studied, from both a four- and five-dimensional perspective.

%\addtolength{\baselineskip}{3mm}   %double-spacing

\newpage

\newsection{Introduction}

In view of the non-linear nature of the Einstein--Maxwell equations, it is somewhat surprising that there should exist a static solution describing an arbitrary superposition of extremal Reissner--Nordstr\"om black holes in neutral equilibrium. This is the well-known Majumdar--Papapetrou solution \cite{Majumdar:1947eu,Papapetrou}. It is actually possible to understand the nature of the balance between the black holes using solely Newtonian arguments: The total force between two (distant) Reissner--Nordstr\"om black holes is given by
\be
F_{\rm int}=-\frac{M_1M_2}{r^2}+\frac{Q_1Q_2}{r^2}\,,
\ee
where the first term is the usual Newtonian gravitational force with a negative sign indicating its attractive nature, while the second term is the electrostatic force. Note that $F_{\rm int}$ will vanish at any distance $r$ if the black holes are extremal, i.e., $M_i=|Q_i|$, and if all the electric charges $Q_i$ have the same sign.

There has been much effort devoted to finding other solutions describing balanced superpositions of black holes. In particular, if the black holes were rotating, there would be an additional gravitational spin-spin interaction \cite{Wald:1972sz} that might be able to balance the other forces present. However, it turns out that the rotating generalisation of the Majumdar--Papapetrou solution, the so-called Perj\'es--Israel--Wilson solution \cite{Perjes:1971gv,Israel:1972vx}, does not describe a superposition of black holes but rather naked singularities \cite{Hartle:1972ya}. Efforts in vacuum Einstein gravity have also been unsuccessful. For example, the double-Kerr solution of Kramer and Neugebauer \cite{Kramer} is now conclusively known not to admit any balanced configurations of black holes (see, e.g., \cite{Neugebauer:2009su,Hennig:2011fp} and references therein).

In fact, up to now, there has been no other solution known to describe a balanced superposition of black holes in either vacuum Einstein or Einstein--Maxwell theory. If we are willing to extend the matter content to include a dilaton (scalar) field---say, in a way motivated by the low-energy effective action of string theory---then generalisations of the Majumdar--Papapetrou solution are known (e.g., \cite{Garfinkle:1990qj,Shiraishi:1992hz,Kallosh:1992ii}). In this case, there will be an extra dilatonic force present, also obeying the inverse-square law at large distances. If we denote the dilaton charge of each black hole by $\Sigma_i$, then the total force between two black holes is
\be
\label{force_eqn}
F_{\rm int}=-\frac{M_1M_2}{r^2}+\frac{Q_1Q_2}{r^2}+\frac{P_1P_2}{r^2}-\frac{\Sigma_1\Sigma_2}{r^2}\,.
\ee
Note that the dilatonic force is attractive when $\Sigma_i$ have the same sign. In the interest of generality, we have also assumed in (\ref{force_eqn}) that each black hole carries a magnetic charge $P_i$, in addition to an electric charge $Q_i$. It was shown in \cite{Garfinkle:1990qj,Shiraishi:1992hz,Kallosh:1992ii} that the total force will vanish if the black holes are extremal, just like those in the Majumdar--Papapetrou solution.

It was pointed out by Kallosh et al.~\cite{Kallosh:1992ii} that the force equation (\ref{force_eqn}) also allows for a purely electric black hole to be in equilibrium with a purely magnetic black hole. In this case, there will not be an electromagnetic force between the two black holes. However, because they now carry opposite dilaton charge, the dilatonic force is repulsive, and is thus able to balance the gravitational force if $M_i=|\Sigma_i|$. Although Kallosh et al.\ were primarily interested in extremal black holes (whose horizons are actually singular in the purely electric or magnetic case), it is conceivable that this force-balance might extend to non-extremal black holes.
Such an electric-magnetic black dihole\footnote{The term `black dihole' has been used by Emparan \cite{Emparan:1999au} to describe a system of two black holes with opposite (dipolar) electric or magnetic charge. In this paper, we shall extend the scope of this term to include the double electric-magnetic black holes considered here.} would represent a class of balanced black holes distinct from the Majumdar--Papapetrou class of solutions.

An explicit example of a balanced electric-magnetic black dihole was constructed by Elvang et al.~\cite{Elvang:2005} (see also \cite{Bates:2003vx,Gaiotto:2005,Bena:2005} for related work), in the context of a four-dimensional supergravity theory obtained by Kaluza--Klein reduction of minimal supergravity in five dimensions. A remarkable feature of this space-time is that it has a non-zero angular momentum arising from the crossed electric and magnetic fields of the two black holes. This phenomenon is already known to arise when an electric charge is placed in the background of a magnetically charged black hole, or vice versa \cite{Garfinkle:1990zx,Kim:2007ca,Bunster:2007sn}.
In fact, such a phenomenon also occurs in classical electrodynamics \cite{Jackson}, when an electric charge is placed a finite distance apart from a magnetic monopole---a system sometimes referred to as Thomson's dipole \cite{Griffiths}.

The Kaluza--Klein origin of the theory considered in \cite{Elvang:2005} means that the electric-magnetic dihole constructed there has a five-dimensional interpretation, which turns out to be no less interesting than the four-dimensional one. As explained in \cite{Elvang:2005}, the magnetic black hole uplifts to a static black hole sitting at the centre of Taub-NUT space, where the nut is located. (In the extremal limit, the magnetic monopole will uplift to the nut of Taub-NUT, with no black hole present.) On the other hand, the electric black hole uplifts to a rotating black ring, with event horizon topology $S^1\times S^2$, surrounding this black hole. Thus the five-dimensional picture consists of what might be called a black Saturn \cite{Elvang:2007rd} on Taub-NUT in five-dimensional minimal supergravity. Equilibrium in this system is achieved as the black ring rotates in the $S^1$ direction at a rate which ensures that the resulting centrifugal force balances the other forces present in the system.

The force-balance condition is often attributed to the presence of supersymmetry, which would be the case for the supersymmetric systems considered in \cite{Kallosh:1992ii} and \cite{Elvang:2005,Gaiotto:2005,Bena:2005,Bates:2003vx}. This argument is even applicable to the original Majumdar--Papapetrou solution, which can be embedded in minimal $N=2$ supergravity \cite{Gibbons:1982fy}. However, non-supersymmetric balanced configurations of black holes do exist too (e.g., \cite{Bena:2009ev}), although in all the known examples the black holes are extremal.

One simple example of this occurs in the Kaluza--Klein reduction of five-dimensional vacuum gravity, which can be regarded in four dimensions as an Einstein--Maxwell-dilaton theory with a particular dilaton coupling. This theory is known to admit two classes of rotating, extremal black holes \cite{Rasheed:1995,Larsen:1999} that carry both electric and magnetic charges. These black holes are not supersymmetric in general \cite{Larsen:1999pu}. Yet, the black holes of one of the classes---the so-called `slowly-rotating' or `ergo-free' class---obeys the force-balance condition. Thus it is possible to construct balanced superpositions of such black holes, at least with $P=Q$ \cite{Chng}.

It was pointed out in \cite{Elvang:2005,Camps:2008hb} that a balanced electric-magnetic black dihole should also exist in this particular Kaluza--Klein theory. The simplest situation would presumably correspond to the case in which both black holes are separately static, in the sense that each black hole is static if the other is removed from the space-time (although, as pointed out above, the space-time will contain a non-zero angular momentum when they are placed a finite distance apart from each other). In the five-dimensional picture, this corresponds to constructing a rotating black ring surrounding a static black hole---in other words, a black Saturn---on Taub-NUT in vacuum gravity. Such a balanced dihole system would be very interesting, because unlike the previously known multi-black-hole systems in which balance can be achieved, it would be expected to be both non-extremal and non-supersymmetric.

Partial progress towards this goal was made by Camps et al.~\cite{Camps:2008hb}, who constructed a rotating black ring on Taub-NUT in five-dimensional vacuum gravity. In the four-dimensional picture, this solution describes a rotating, electrically charged black hole a finite distance apart from a magnetic monopole. The rotation of the black hole arises from the fact that the black ring is rotating in both possible directions in five dimensions, even though there is only one rotational parameter governing the solution. This means that the angular momentum and electric charge of the black hole are not independent parameters.

In \cite{Chen:2012zb}, a simpler version of the above solution was constructed by the present authors, in which the black ring on Taub-NUT rotates purely in the $S^1$ direction, without any rotation in the $S^2$ direction. For this reason, it is appropriate to refer to it as an Emparan--Reall black ring \cite{Emparan:2001} on Taub-NUT. In the four-dimensional picture, this solution describes a static, electrically charged black hole a finite distance apart from a magnetic monopole. Although balance is possible in this system, it suffers from the drawback that the magnetic monopole is actually a singular object (even though it is completely regular from the five-dimensional viewpoint). Fortunately, this can be remedied in principle by lifting the magnetic monopole above extremality to become a magnetic black hole.

It is the goal of the present paper to carry this out in practice. It turns out to be easier to construct the five-dimensional version of this solution, which corresponds to introducing a static black hole at the centre of the above black ring. This is because a powerful solution-generating technique, known as the inverse-scattering method (ISM) \cite{Belinski:2001,Pomeransky:2005sj}, is applicable to vacuum space-times with sufficient symmetry. In particular, it has been very successfully used to construct various black-ring and related solutions in five dimensions, such as the doubly rotating black ring \cite{Pomeransky:2006,Chen:2011jb}, black Saturn \cite{Elvang:2007rd}, and double-black-ring solutions \cite{Iguchi:2007is,Evslin:2007fv,Izumi:2007qx,Elvang:2007hs}. It was also the method used to construct the rotating black rings on Taub-NUT in \cite{Chen:2012zb}. The ISM construction of the black Saturn on Taub-NUT will be briefly described in Sec.~2.

Having obtained the five-dimensional solution, we then calculate its rod structure to verify that it describes a black Saturn system. This is the subject of Sec.~3. The rod structure analysis will also yield a condition for the configuration to be balanced. We also present the four-dimensional form of this solution and its corresponding rod structure. The latter will confirm that this solution describes a double-black-hole, or dihole, system.

The physical properties of the solution are then discussed in Sec.~4, from both a five- and four-dimensional perspective. In particular, a proof is given (with details in Appendix A) that when the balance condition is imposed, both the five- and four-dimensional space-times are regular and well-behaved on and outside the event horizons. Focusing on the four-dimensional solution, we then carry out a study of the parameter and phase spaces of the electric-magnetic dihole system when balance is imposed. One of the results of this study is that there is a high degree of non-uniqueness in the phase space. It turns out that there is, in general, a continuous infinity of balanced dihole configurations with the same conserved charges.

In Sec.~5, the forces between the two four-dimensional black holes are studied in the far-separation approximation. When balance is imposed, it is confirmed that there is, as expected, a repulsive dilatonic force balancing the gravitational force by (\ref{force_eqn}), so that the system is in neutral equilibrium to leading order. Then in Sec.~6, we show how various known limits of this solution can be recovered, in complete consistency with the above five- and four-dimensional interpretations of the solution. The paper ends with a brief discussion of some open questions and future directions in Sec.~7.

\newsection{ISM construction}

\begin{figure}[t]
\begin{center}
\includegraphics{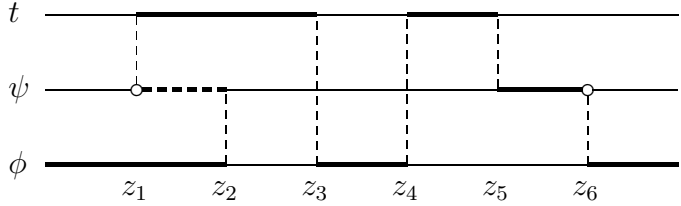}
\caption{The rod sources of the seed solution for the black Saturn on Taub-NUT. The thin lines denote the $z$-axis and the thick lines denote rod sources of mass $\frac{1}{2}$ per unit length along this axis. The dashed horizontal line denotes a rod source with negative mass density $-\frac{1}{2}$. Small circles represent the operations of removing solitons from the seed, each with a BZ vector having a single non-vanishing component along the coordinate that labels the $z$-axis where the circle is placed.}
\label{fig_bs_tn}
\end{center}
\end{figure}

To construct the black Saturn on Taub-NUT, we start with a seed solution having the rod structure as shown in Fig.~\ref{fig_bs_tn}. In Weyl--Papapetrou coordinates \cite{Emparan:2001wk}
\be
\dif s^2=G_{ab}\,\dif x^a\dif x^b+\me^{2\gamma}(\dif\rho^2+\dif z^2)\,,
\ee
the explicit solution corresponding to this rod structure can be directly read off as
\ba
G_0&=&{\textrm{diag}}\, \bigg\{ -{\frac {{\mu_1\mu_4}}{{\mu_3\mu_5}}},\frac{\mu_2\mu_5}{\mu_1\mu_6}, \frac{\rho^2\mu_3\mu_6}{\mu_2\mu_4}\bigg\}\,,\\
\me^{2\gamma_0}&=&k^2\frac{\mu_3 \mu_6 R_{13}  R_{45} R_{12} R_{56} R_{23}  R_{46} R_{34}^2R_{26}^2 R_{15}^2}{\mu_2\mu_4 R_{11} R_{22} R_{33} R_{44}R_{55} R_{66} R_{14} R_{35} R_{25}R_{16}R_{36}R_{24} }\,,
\ea
where $\mu_i\equiv\sqrt{\rho^2+(z-z_i)^2}-(z-z_i)$, $R_{ij}\equiv\rho^2+\mu_i\mu_j$, and $k$ is an arbitrary integration constant which will be set to be one to ensure that the reduced solution in four dimensions is asymptotically flat. This seed differs from that used in \cite{Chen:2012zb} in that there is an extra rod placed between $z_4$ and $z_5$ in Fig.~\ref{fig_bs_tn}. Using the ISM, we then perform the following soliton transformations on the above seed:

\begin{enumerate}
  \item Remove a soliton at each of $z_1$ and $z_6$, with trivial Belinski--Zakharov (BZ) vector $(0,1,0)$;
  \item Add back a soliton at each of $z_1$ and $z_6$, with non-trivial  BZ vectors $(4z_1^2C_1,1,0)$ and $(0,1,8z_6^3C_2)$ respectively. Here, $C_1$ and $C_2$ are the new, so-called BZ parameters.
\end{enumerate}

The above construction is rather similar to that of \cite{Chen:2012zb}, to which we refer the reader for some technical details. Here we briefly describe some key points of the construction. After the first step, we obtain the new $G$-matrix:
\be
\tilde{G}_0= {\textrm{diag}}\,  \bigg\{-\frac{\mu_1 \mu_4}{\mu_3\mu_5} ,{\frac {{\mu_1}{\mu_2}{\mu_5}\mu_6}{{\rho}
^{4}}},{\frac {\rho^2{\mu_3}\mu_6}{\mu_2\mu_4}} \bigg\}\,.
\ee
The corresponding generating matrix $\tilde{\Psi}_0$ can be obtained directly by performing the following replacements to the $\tilde{G}_0$ matrix: $\mu_i\rightarrow \mu_i-\lambda$ and $\rho^2\rightarrow \rho^2-2z\lambda-\lambda^2$, where $\lambda$ is a spectral parameter. We mention that the identity $\mu_i^2+2z\mu_i-\rho^2=2z_i\mu_i$ has been used to simplify the vectors $m^{(k)}$ involved in the construction. The solution thus generated can be written in a compact form when collected in terms of the BZ parameters $C_1$ and $C_2$, which will be presented and analysed in the following section.

\newsection{The metric and rod structure}

The five-dimensional metric we obtained can be written in Weyl--Papapetrou coordinates in the following form:
\ba
\label{metric_bs_tn}
\dif s^2&=&\frac{\mu_1\mu_2\mu_5\mu_6 K}{H}\left(\dif \psi+\omega_1\, \dif t+\omega_2\,\dif \phi\right)^2-\frac{\mu_4 F}{\mu_1\mu_3\mu_5 K}\left(\dif t+\omega_3 \,\dif \phi\right)^2+\frac{\mu_3\rho^2 H}{\mu_2\mu_4\mu_6 F}\,\dif \phi^2\cr
&&+RH\left(\dif \rho^2+\dif z^2\right),
\ea
where
\ba
\label{functions_A}
\omega_1&=&\frac{C_1 R_{11} \sqrt{\mu_4}}{\mu_1\mu_5\sqrt{\mu_2\mu_3\mu_6} K}\left(C_2^2\rho^4\sqrt{M_1M_3}+\mu_6\sqrt{M_2M_4}\right),\cr
\omega_2&=&\frac{C_2\rho^2R_{66}\sqrt{\mu_3}}{\mu_2\mu_6\sqrt{\mu_1\mu_4\mu_5}K}
\left(C_1^2\rho^2\sqrt{M_1M_2}+\mu_1\sqrt{M_3M_4}\right),\cr
\omega_3&=&\frac{C_1C_2\mu_3\rho^2R_{11}R_{66}\sqrt{\mu_1\mu_5 M_1 M_4}}{\mu_4\mu_{61}\sqrt{\mu_2\mu_6}F}\,,
\ea
and the functions $R$, $H$, $F$, $K$ and $M_{1,2,3,4}$ are defined as
\ba
\label{functions_HF}
R&=&\frac{\mu_3R_{12}R_{13}R_{23}R_{45}R_{46}R_{56}R_{15}^2R_{26}^2R_{34}^2}{\mu_1^2\mu_2\mu_4\mu_6M_4R_{11}R_{22}R_{33}R_{44}R_{55}R_{66}R_{14}R_{16}R_{24}R_{25}R_{35}R_{36}}\,,\cr
H&=&C_1^2C_2^2\rho^8M_1+C_1^2\mu_6^2\rho^2 M_2-C_2^2\mu_1^2\rho^4M_3+\mu_1^2\mu_6^2 M_4\,,\cr
F&=&-C_1^2C_2^2\rho^6M_1+C_1^2\rho^2M_2+C_2^2\mu_1^2\rho^2M_3+\mu_1^2M_4\,,\cr
K&=&C_1^2C_2^2\rho^4M_1-C_1^2M_2+C_2^2\rho^2M_3+M_4\,,\cr
M_1&=&\mu_3\mu_4\mu_{21}^2\mu_{31}^2\mu_{51}^4\mu_{61}^2\mu_{62}^4\mu_{64}^2\mu_{65}^2R_{16}^2\,,\cr
M_2&=&\mu_1\mu_2^2\mu_4^2\mu_5\mu_6^4\mu_{21}^2\mu_{31}^2\mu_{51}^4\mu_{63}^2R_{16}^2\,,\cr
M_3&=&\mu_1^2\mu_2\mu_{3}^2\mu_{5}^2\mu_{6}\mu_{41}^2\mu_{62}^4\mu_{64}^2\mu_{65}^2R_{16}^2\,,\cr
M_4&=&\mu_1^3\mu_{2}^3\mu_{3}\mu_{4}\mu_{5}^3\mu_{6}^5\mu_{41}^2\mu_{63}^2\,,
\ea
with $\mu_{ij}\equiv \mu_i-\mu_j$.

In this paper, we impose the ordering $z_i>z_j$ if $i>j$. It is then not difficult to show that the inequality $\mu_i>\mu_j>0$ holds if $i>j$, which in turn implies that $\mu_{ij}>0$. The rod structure of the above solution can be calculated directly, following the prescription of \cite{Chen:2010zu,Chen:2010ih}. It consists of six turning points $z_1,...,z_6$, dividing the $z$-axis into seven rods. Counting the rods from the left, we then look for the conditions to join up Rods 1 and 2, as well as Rods 6 and 7, i.e., the conditions to ensure that Rod 1 has the same (normalised) direction as Rod 2, and Rod 6 has the same direction as Rod 7. These operations are very similar to those done in \cite{Chen:2012zb}, with the motivations explained therein. This gives equations for $C_1^2$ and $C_2^2$; without loss of generality, we choose the following solutions for them:
\be
\label{C12}
C_1=-\sqrt{\frac{z_{41}}{16z_{21}z_{31}z_{51}^2z_{61}}}\,,\qquad C_2=\sqrt{\frac{z_{63}}{16z_{61}z_{62}^2z_{64}z_{65}}}\,,
\ee
where $z_{ij}\equiv z_i-z_j$.
This effectively removes the turning points $z_1$ and $z_6$ from the rod structure. From now on, we {\it always\/} assume that the above values of $C_{1,2}$ are imposed unless otherwise specified, hence the solution is completely characterised by the positions of the turning points $z_1,...,z_6$ (including the so-called phantom points at $z_1$ and $z_6$). 

\begin{figure}[t]
\begin{center}
\includegraphics{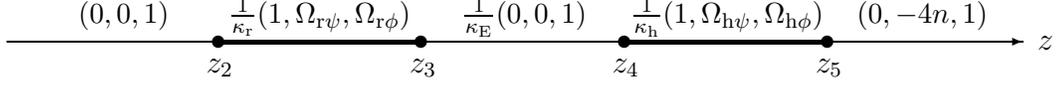}
\caption{The rod structure of the black Saturn on Taub-NUT, where $n$ is the NUT-charge parameter. The thick lines denote the locations of the horizons, while the thin lines denote the locations of the axes. The system is balanced in the case $\kappa_{\rm E}=1$.}
\label{Figure_BS_TN}
\end{center}
\end{figure}

Thus, the rod structure of the final solution has only four genuine turning points $z_2,...,z_5$. They divide the $z$-axis into five rods as follows:
\begin{itemize}
\item Rod I: a semi-infinite space-like rod located at $(\rho=0, z\leq z_2)$, with direction $\ell_{\rm I}=(0,0,1)$.

\item Rod II: a finite time-like rod located at $(\rho=0, z_2\leq z\leq z_3)$, with direction $\ell_{\rm II}=\frac{1}{\kappa_{\rm r}}(1,\Omega_{{\rm r}\psi},\Omega_{{\rm r}\phi})$, where the surface gravity $\kappa_{\rm r}$ and angular velocities $\Omega_{{\rm r}\psi}$ and $\Omega_{{\rm r}\phi}$ are given by
\ba
\label{kappa_Omega_r}
\kappa_{\rm r}=\sqrt{\frac{z_{41}z_{42}z_{52}z_{61}}{4z_{31}z_{32}z_{51}^2z_{62}^2}}\,,\qquad
\Omega_{{\rm r}\psi}=\sqrt{\frac{z_{21}z_{41}z_{61}}{z_{31}z_{51}^2}}\,,\qquad
\Omega_{{\rm r}\phi}=-\sqrt{\frac{z_{21}z_{41}z_{64}z_{65}}{4z_{31}z_{51}^2z_{62}^2z_{63}}}\,.
\ea

\item Rod III: a finite space-like rod located at $(\rho=0, z_3\leq z\leq z_4)$, with direction $\ell_{\rm III}=\frac{1}{\kappa_{\rm E}}(0,0,1)$, where the Euclidean surface gravity $\kappa_{\rm E}$ is defined as
\be
\label{kappa_E}
{\kappa_{\rm E}}=\frac{\sqrt{z_{41}z_{42}z_{52}z_{53}z_{61}z_{63}}}{z_{43}z_{51}z_{62}}\,.
\ee

\item Rod IV: a finite time-like rod located at $(\rho=0, z_4\leq z\leq z_5)$, with direction $\ell_{\rm IV}=\frac{1}{\kappa_{\rm h}}(1,\Omega_{{\rm h}\psi},\Omega_{{\rm h}\phi})$, where the surface gravity $\kappa_{\rm h}$ and angular velocities $\Omega_{{\rm h}\psi}$ and $\Omega_{{\rm h}\phi}$ are given by
\ba
\label{kappa_Omega_h}
\kappa_{\rm h}=\sqrt{\frac{z_{52}z_{53}z_{61}z_{63}}{4z_{54}z_{64}z_{51}^2z_{62}^2}}\,,\qquad
\Omega_{{\rm h}\psi}=\sqrt{\frac{z_{21}z_{31}z_{61}}{z_{41}z_{51}^2}}\,,\qquad
\Omega_{{\rm h}\phi}=-\sqrt{\frac{z_{21}z_{31}z_{63}z_{65}}{4z_{41}z_{51}^2z_{62}^2z_{64}}}\,.
\ea

\item Rod V: a semi-infinite space-like rod located at $(\rho=0, z\geq z_5)$, with direction $\ell_{\rm V}=(0,-4n,1)$, where the NUT charge $n$ is given by
\be
\label{n}
n=\sqrt{\frac{z_{62}^2z_{64}z_{65}}{4z_{61}z_{63}}}\,.
\ee
\end{itemize}
This rod structure is illustrated in Fig.~\ref{Figure_BS_TN}.

Here we identify \{$\ell_{\rm I}$, $\ell_{\rm V}$\} as the two independent $2\pi$-periodic generators of the $U(1)\times U(1)$ isometry subgroup of this solution, thus making the following identifications on the coordinates \cite{Chen:2010zu}:
\be
\label{identifications_coor}
(\psi,\phi)\rightarrow (\psi,\phi+2\pi)\,, \qquad (\psi,\phi)\rightarrow (\psi-8n\pi,\phi+2\pi)\,.
\ee

Recall that there are two time-like rods, II and IV, in the rod structure. It can be seen from the above identifications that they represent Killing horizons with topologies $S^1\times S^2$ and $S^3$ respectively, corresponding to a black ring and a black hole (hence the use of the subscripts `r' and `h' in (\ref{kappa_Omega_r}) and (\ref{kappa_Omega_h}) respectively). Hence the solution describes what is known as a black Saturn; moreover, we will see in the following section that the asymptotic geometry of the space-time is of Taub-NUT type. Under the above identifications, the solution (\ref{metric_bs_tn}) in general possesses a conical singularity along Rod III. The excess angle $\Delta\phi$ along the axis represented by this rod can then be calculated to be
\be
\label{Delta_phi}
\Delta\phi=2\pi(\kappa_\mathrm{E}-1)\,.
\ee
A completely regular class of space-times, describing a balanced black Saturn on Taub-NUT, can be obtained by imposing $\ell_{\rm I}=\ell_{\rm III}$, or equivalently,
\be
\label{balance_condition}
\kappa_{\rm E}=1\,.
\ee

Note that the solution (\ref{metric_bs_tn}) has six parameters $z_1,...,z_6$, of which only five are non-trivial, since there is a translational symmetry along the $z$ direction. So only the relative differences between the $z_i$'s are physical, which is already manifest in the quantities appearing in the rod structure. They can be interpreted as follows: Roughly speaking, $z_{21}$, $z_{32}$ and $z_{43}$ respectively characterise the angular momentum, mass and radius of the black ring, $z_{54}$ characterises the mass of the black hole, and $z_{65}$ determines the NUT charge of the space-time. It is clear that after imposing the balance condition, we are left with one fewer independent parameter.

Now performing dimensional reduction on (\ref{metric_bs_tn}) along the direction $\frac{\partial}{\partial \psi}$, we obtain a solution of four-dimensional Kaluza--Klein theory. The resulting metric takes the following form:
\be
\label{metric_4D}
\dif s^2=\bigg(\frac{\mu_1\mu_2\mu_5\mu_6 K}{H}\bigg)^{\frac{1}{2}}\bigg[-\frac{\mu_4 F}{\mu_1\mu_3\mu_5 K}\left(\dif t+\omega_3 \,\dif \phi\right)^2+\frac{\mu_3\rho^2 H}{\mu_2\mu_4\mu_6 F}\,\dif \phi^2+RH\big(\dif \rho^2+\dif z^2\big)\bigg]\,,
\ee
where the functions $\omega_3$, $R$, $H$, $F$ and $K$ are given in (\ref{functions_A}) and (\ref{functions_HF}). The gauge potential $A$ is
\be
A=\hbox{$\frac{1}{2}$}(\omega_1\,\dif t+\omega_2\,\dif\phi)\,,
\ee
where $\omega_{1,2}$ are defined as in (\ref{functions_A}), and the dilaton field $\sigma$ is given by
\be
\label{dilaton}
{\rm e}^{\sigma}=\bigg(\frac{\mu_1\mu_2\mu_5\mu_6 K}{H}\bigg)^{\frac{\sqrt3}{4}}.
\ee

\begin{figure}[t]
\begin{center}
\includegraphics{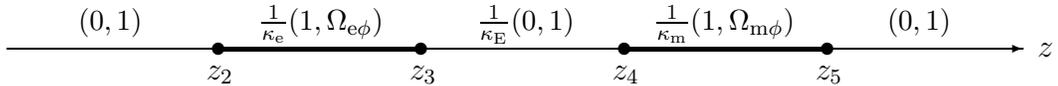}
\caption{The rod structure of the reduced four-dimensional space-time, describing an electric-magnetic dihole in an asymptotically flat space-time. The system is balanced in the case $\kappa_{\rm E}=1$.}
\label{fig_4D_rod}
\end{center}
\end{figure}

The rod structure of the four-dimensional metric (\ref{metric_4D}) can also be readily calculated. We note that it is actually the same as that of (\ref{metric_bs_tn}), but now with the $\psi$-components of all the rod directions removed. This rod structure is shown schematically in Fig.~\ref{fig_4D_rod}, and it clearly describes a configuration of two black holes, i.e., a dihole. As we will see in the following section, the black hole represented by Rod II is electrically charged, while that represented by Rod IV is magnetically charged. Hence, in Fig.~\ref{fig_4D_rod}, we have renamed the subscripts `r' and `h' more appropriately by `e' and `m' respectively, but the values of the surface gravities and angular velocities are understood to be still given by (\ref{kappa_Omega_r}) and (\ref{kappa_Omega_h}).

In the four-dimensional picture, the five independent parameters can be interpreted as follows: Roughly speaking, $z_{21}$ and $z_{65}$ characterise the electric and magnetic charges carried by the two black holes respectively, $z_{32}$ and $z_{54}$ characterise their individual irreducible masses, and $z_{43}$ determines the distance between them. As in the five-dimensional picture, balance can be achieved by imposing the condition (\ref{balance_condition}), in which case we are left with four independent parameters.

\newsection{Physical properties}

We begin by examining the asymptotic geometry of the black-Saturn solution (\ref{metric_bs_tn}). Since infinity of the space-time is located at $\rho,z\rightarrow \infty$, we define
\be
\rho=r\sin\theta\,,\qquad z=r\cos\theta\,,
\ee
and set $r\rightarrow \infty$. The metric (\ref{metric_bs_tn}) then becomes the direct product of a flat time dimension and the Taub-NUT geometry, i.e.,
\be
\label{asymptotic_structure}
\dif s^2\rightarrow -\dif t^2+\left[\dif \psi+2n(1+\cos\theta)\,\dif \phi\right]^2+\dif r^2+r^2(\dif \theta^2+\sin^2\theta\,\dif \phi^2)\,,
\ee
where $n$ is given by (\ref{n}). Thus, (\ref{metric_bs_tn}) indeed describes a black Saturn on Taub-NUT.

An interesting property of this solution is that the black ring does not carry any $S^2$-rotation, i.e., the Komar angular momentum of the black ring associated to the Killing vector $\ell_{\rm I}=\frac{\partial}{\partial\phi}$ vanishes:
\be
J_{{\rm r}\phi}=0\,.
\ee
A similar situation occurs in the case of the singly rotating black ring on Taub-NUT constructed in \cite{Chen:2012zb}. Moreover, the black hole in the present system does not carry any rotations at all, associated to either the direction $\frac{\partial}{\partial \psi}$ or the direction $\frac{\partial}{\partial \phi}$. Hence the solution (\ref{metric_bs_tn}) describes the simplest possible rotating black Saturn on Taub-NUT: an Emparan--Reall black ring surrounding a static black hole placed at the NUT. However, we note that, even though the black hole carries no intrinsic angular momenta, its angular velocities $\Omega_{{\rm h}\psi}$ and $\Omega_{{\rm h}\phi}$ are both non-zero due to frame-dragging effects caused by the rotation of the black ring in the non-trivial Taub-NUT background space-time. A similar phenomenon has also been observed in the black-Saturn system in asymptotically flat space-time \cite{Elvang:2007rd}.

It can be shown that the five-dimensional space-time described by the metric (\ref{metric_bs_tn}), as well as the four-dimensional reduced metric (\ref{metric_4D}), is regular and well-behaved on and outside the event horizons if the balance condition (\ref{balance_condition}) is imposed. In particular, this means that the space-time is free of naked singularities and closed time-like curves (CTCs). This can be shown, for example, along the lines of the smoothness proof of \cite{Chrusciel:2010ix,Chrusciel:2010nz}, and we give the key points of this proof in Appendix A.

It can also be shown that ergoregions exist in both the five- and four-dimensional space-times. In the five-dimensional case, this follows from the fact that the $g_{tt}$ component of the metric vanishes at the phantom point $z_1$, so the ergo-surface will pass through at least this point along the $z$-axis. On the other hand, it can be checked that the $g_{tt}$ component of the four-dimensional metric vanishes at the turning points $z_2,...,z_5$. This means that the two ergo-surfaces will touch the event horizons of the two black holes at these points, consistent with Hajicek's theorem \cite{Hajicek}.

In Kaluza--Klein theory, the solution (\ref{metric_4D})--(\ref{dilaton}) describes an electric-magnetic dihole system in an asymptotically flat space-time. The total mass, angular momentum and scalar charge of the whole system, and the electric and magnetic charges of each black hole are calculated to be
\ba
M_{\rm tot}=\frac{z_{54}+z_{64}+z_{31}+z_{32}}{4}\,,\quad J_{\rm tot}=-\sqrt{\frac{z_{21}z_{31}z_{64}z_{65}z_{51}^2z_{62}^2}{16z_{41}z_{63}z_{61}^2}}\,,\quad \Sigma_{\rm tot}=\frac{\sqrt{3}(z_{21}-z_{65})}{4}\,,\cr
Q_{\rm e}=\sqrt{\frac{z_{21}z_{31}z_{51}^2}{4z_{41}z_{61}}}\,,\qquad P_{\rm e}=0\,,\qquad Q_{\rm m}=0\,,\qquad P_{\rm m}=-\sqrt{\frac{z_{64}z_{65}z_{62}^2}{4z_{61}z_{63}}}\,.\hspace{0.6in}
\ea
In particular, we see that each of the two black holes in Kaluza--Klein theory is either purely electric or purely magnetic; in other words, the electric black hole carries no magnetic charge and vice versa. Also note that the above quantities are not independent; a simple relation between them is
\be
J_{\rm tot}=P_{\rm m}Q_{\rm e}\,.
\ee
This relation is a direct consequence of the fact that from the five-dimensional viewpoint, neither the black ring nor the black hole carries an $S^2$-rotation.\footnote{This can be more easily seen if we use the coordinates ($\psi'$,$\phi'$) which are linear combinations of ($\psi$,$\phi$), such that $\ell_{\rm I}$ and $\ell_{\rm V}$ can now be expressed as $(0,2n,1)$ and $(0,-2n,1)$ respectively. In the case when the black hole is made to vanish, i.e., $z_4=z_5$, a detailed discussion can be found in \cite{Chen:2012zb}.} For completeness, we also calculate the areas of the two black-hole event horizons, which are respectively
\be
A_{\rm e}=\sqrt{\frac{16\pi^2 z_{31}z_{32}^3z_{51}^2z_{62}^2}{z_{41}z_{42}z_{52}z_{61}}}\,,\qquad A_{\rm m}=\sqrt{\frac{16\pi^2 z_{51}^2z_{54}^3z_{62}^2z_{64}}{z_{52}z_{53}z_{61}z_{63}}}\,.
\ee

We define the `irreducible mass' of each black hole to be its mass when its charge is turned off and the other black hole made to vanish. In the current case, the irreducible masses of the electric and magnetic black holes are respectively
\be
M_{\rm e}^{\rm irr}=\frac{z_{32}}{2}\,,\qquad M_{\rm m}^{\rm irr}=\frac{z_{54}}{2}\,.
\ee

Now we turn to a study of the parameter and phase spaces of the electric-magnetic dihole system, when balance is imposed. To simplify the analysis, we will use the total irreducible mass of the system as a scale parameter. We now write the parameters of the solution in units of $z_{32}+z_{54}$ as follows:
\be
\label{parameterisation}
z_{21}=a\,,\qquad z_{32}=1-m\,,\qquad z_{43}=l\,,\qquad z_{54}=m\,,\qquad z_{65}=b\,.
\ee
In this way, all the physical quantities calculated in this paper can be regarded as having units of some integer powers of $z_{32}+z_{54}$. In this parameterisation, the balance condition (\ref{balance_condition}) reduces to
\be
\label{parameter_constraint}
{\frac { ( l+1 ) ( l+m )( l+1-m )( a+l+1-m ) ( b+l+m )  ( a+b+l+1
 )    }{{l}^{2} ( a+l+1
 ) ^{2} ( b+l+1 ) ^{2}}}=1\,.
\ee
The parameter space is then a surface given by (\ref{parameter_constraint}) in the four-dimensional space $(a,b,l,m)$, with $a,b,l>0$ and $0\leq m\leq1$. Note that $m=\frac{1}{2}$ corresponds to the case in which the two black holes have the same irreducible mass, while $m=0$ (1) corresponds to the case in which the magnetic (electric) black hole becomes extremal.

\begin{figure}[!ht]
\begin{center}
\includegraphics[width=3in]{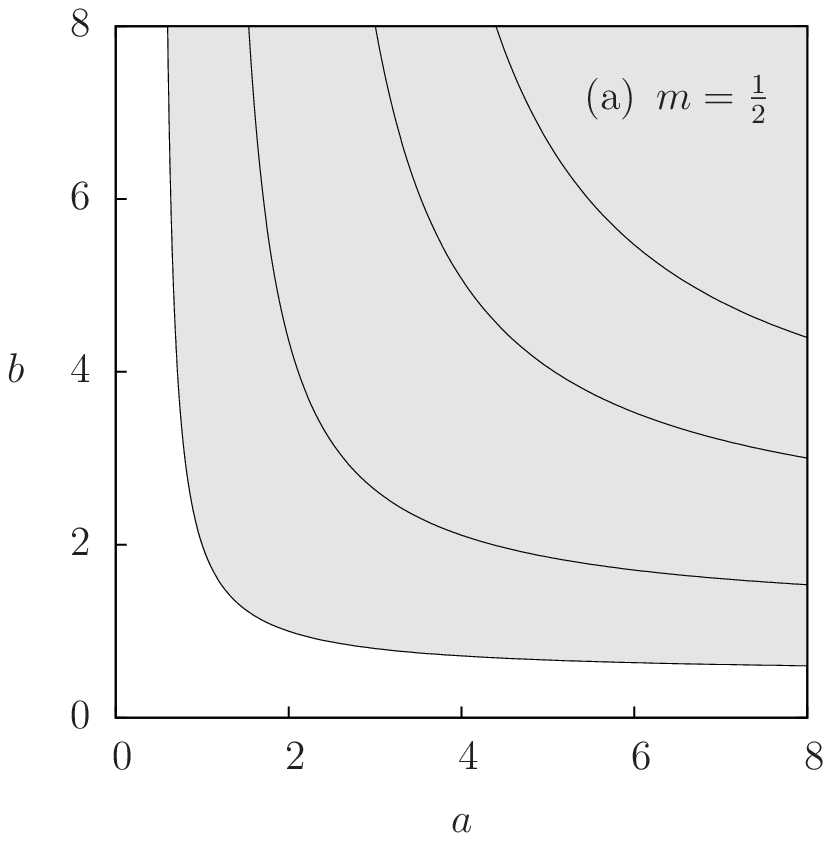}
\includegraphics[width=3in]{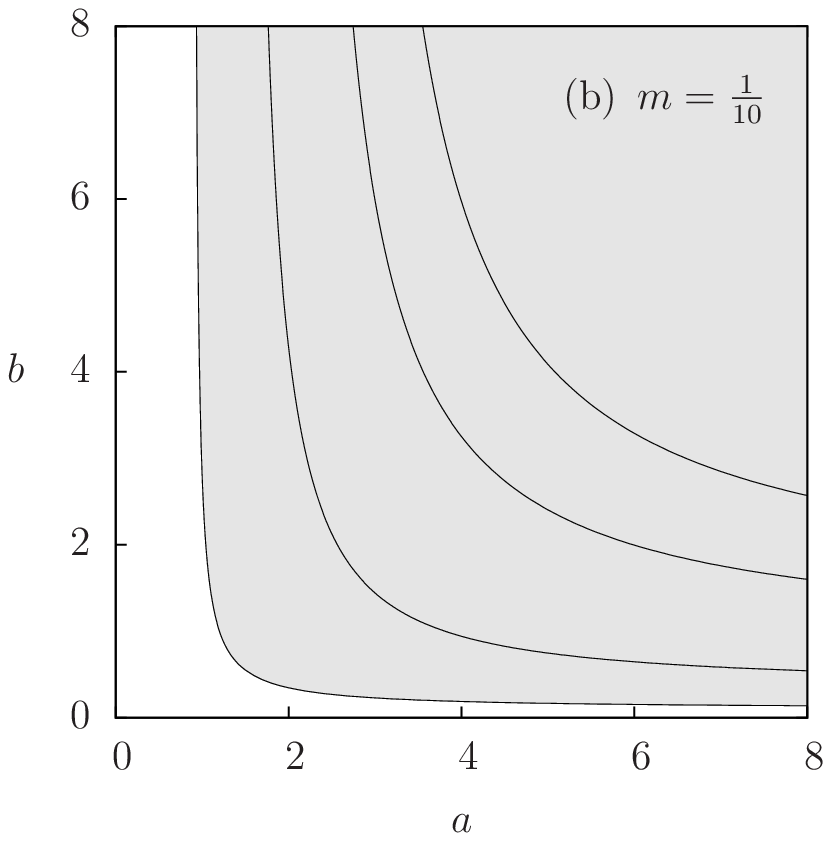}
\includegraphics[width=3in]{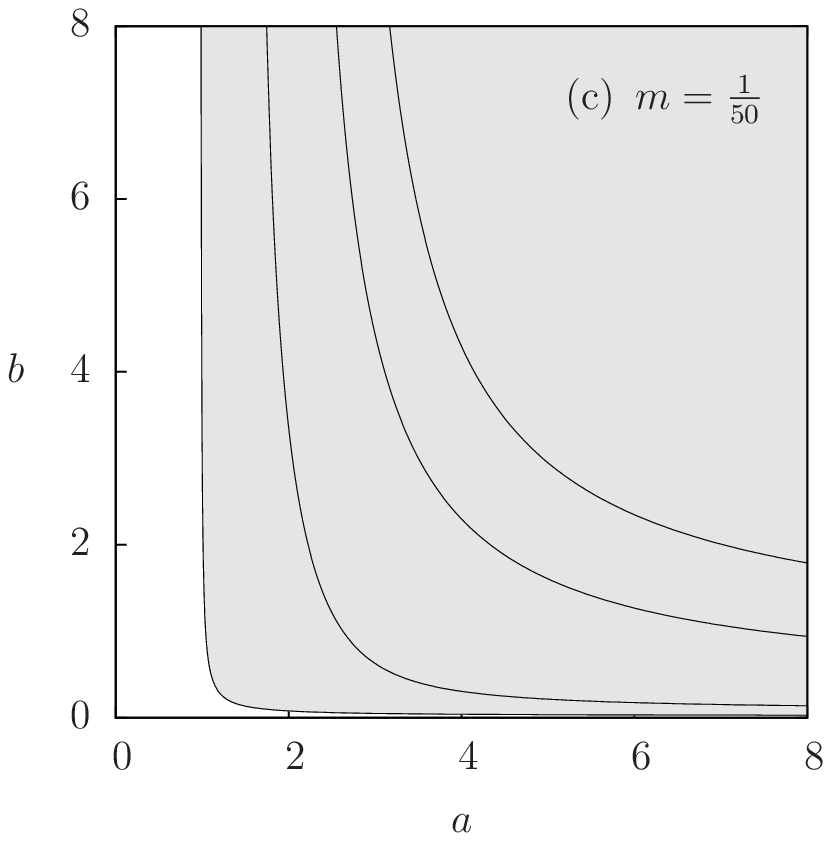}
\includegraphics[width=3in]{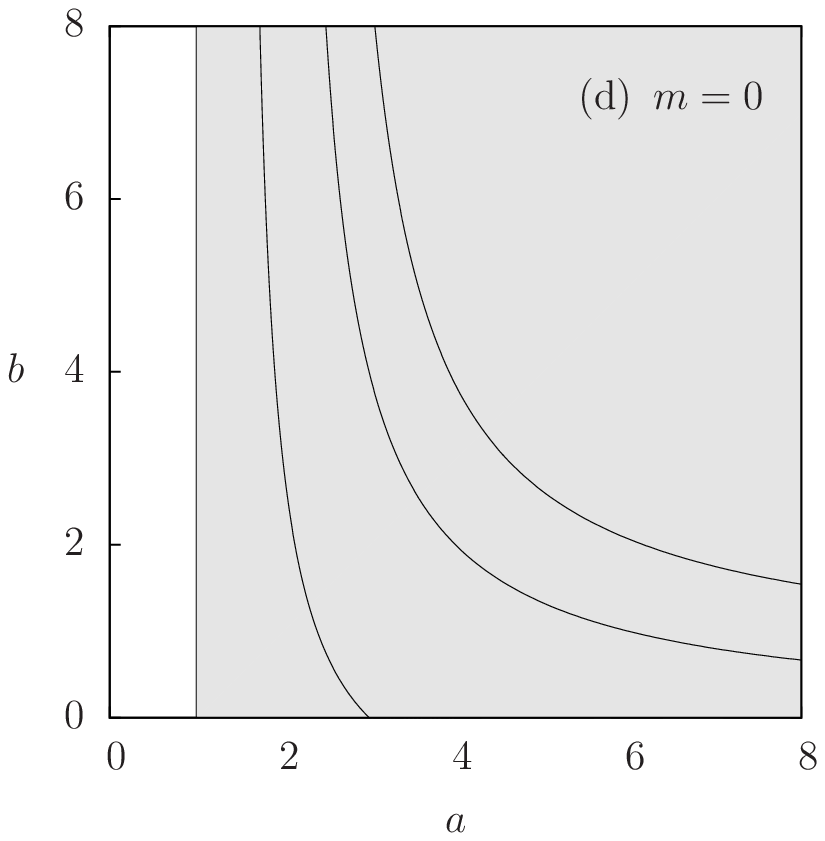}
\caption{The parameter space of the balanced dihole system for the cases (a)~$m=\frac{1}{2}$, (b)~$m=\frac{1}{10}$, (c)~$m=\frac{1}{50}$, and (d)~$m=0$. In each case, four representative $b$ versus $a$ curves for fixed $l$ are plotted; the values of $l$ for these curves from upper right to lower left are respectively $\frac{3}{4}$, $1$, $2$ and $\infty$. The parameter space is the union of all these curves with fixed $l$ taking values from $0$ to $\infty$, and is represented by the shaded area in each plot.}
\label{fig_parameter_space}
\end{center}
\end{figure}

Note that the balance condition in the form (\ref{parameter_constraint}) is a quadratic equation in either of the variables $a$ or $b$. So we can directly solve say $b$ in terms of $a$, $l$ and $m$. We have found that, given a fixed $l$ and $m$, for (\ref{parameter_constraint}) to have valid solutions, $a$ must exceed a minimum given by
\ba
a_{\rm min}&\equiv&\frac{1}{2{l}^{2}}\Big\{ -{l}^{3}+l+ml-{m}^{2}l+m-{m}^{2}+ \big[ ( l+1 )( l+m )( l+1-m )\cr
&&\qquad\times  ( {l}^{3}+2{l}^{2}-4m{l}^{2}+l+ml-{m}^{2}l+m-{m}^{2} ) \big]^{\frac{1}{2}}\Big\}\,.
\ea
If $a>a_{\rm min}$, $b$ has exactly one solution which we denote by $b=b_1(l,m,a)$; if instead $0<a<a_{\rm min}$, $b$ has no valid solutions, and so the system cannot be balanced. Similarly, if $b>b_{\rm min}$, where 
\ba
b_{\rm min}&\equiv&\frac{1}{2{l}^{2}}\Big\{ -{l}^{3}+l+ml-{m}^{2}l+m-{m}^{2}+ \big[ ( l+1 )( l+m )( l+1-m )\cr
&&\qquad\times  ( {l}^{3}-2{l}^{2}+4m{l}^{2}+l+ml-{m}^{2}l+m-{m}^{2} ) \big]^{\frac{1}{2}}\Big\}\,,
\ea
$a$ has exactly one solution $a=a_1(l,m,b)$; otherwise, the solution cannot be balanced. The above formulae immediately imply that $a$ and $b$ have an infimum (for fixed $m$ and all possible $l\in (0,\infty)$):
\be
a_{\rm MIN}\equiv1-m\,,\qquad b_{\rm MIN}\equiv m\,,
\ee
so if either $0<a\leq a_{\rm MIN}$ (i.e., $0<z_{21}\leq z_{32}$) or $0<b\leq b_{\rm MIN}$ (i.e., $0<z_{65}\leq z_{54}$), the balance condition cannot be satisfied for any value of $l$. We plot the curves of the function $b=b_1(l,m,a)$ in Fig.~\ref{fig_parameter_space}(a) for the case $m=\frac{1}{2}$ and several representative values of $l$. The parameter space for this value of $m$ is the union of all such curves with fixed $l$ taking values from $0$ to $\infty$. In the case $l\rightarrow \infty$, the balance condition simply becomes
\be
(a-a_{\rm MIN})(b-b_{\rm MIN})=3m(1-m)\,.
\ee
The corresponding curve is shown as the lowest left one in Fig.~\ref{fig_parameter_space}(a), which forms the boundary of the whole parameter space. 

We have also plotted in Figs.~\ref{fig_parameter_space}(b), (c) and (d) the parameter space for the cases $m=\frac{1}{10}$, $\frac{1}{50}$ and $0$, respectively. The first two cases are qualitatively similar to the $m=\frac{1}{2}$ case, except that they are offset to the lower-right of the graph. However, the $m=0$ parameter space in Fig.~\ref{fig_parameter_space}(d) exhibits some qualitatively different features. In this case, it can be checked that constant $l$ curves with $l>1$ will intersect the $a$-axis. In particular, the $l\rightarrow \infty$ curve is the vertical line $a=1$, which forms the boundary of the parameter space. This $m=0$ case actually corresponds, from the five-dimensional viewpoint, to the balanced Emparan--Reall black ring on Taub-NUT found in \cite{Chen:2012zb} (see Sec.~6.1). The fact that constant $l<1$ curves do not intersect the $a$-axis is consistent with the results of \cite{Chen:2012zb}.

Although we have not explicitly plotted the parameter space for any case in which $m>\frac{1}{2}$, we note that this case can be obtained from the $m<\frac{1}{2}$ case by the symmetry of (\ref{parameter_constraint}) under the interchange $m\leftrightarrow 1-m$ and $a\leftrightarrow b$. Thus, the cases $m=\frac{9}{10}$, $\frac{49}{50}$ and $1$ can be simply obtained from Figs.~\ref{fig_parameter_space}(b), (c) and (d) respectively by interchanging the $a$- and $b$-axes. In particular, the case $m=1$ corresponds, from the five-dimensional viewpoint, to an extremal rotating black ring around a static black hole on Taub-NUT. In parallel with the $m=0$ case discussed above, we can infer that constant $l$ curves with $l>1$ will intersect the $b$-axis.

\begin{figure}[!ht]
\begin{center}
\includegraphics[width=3in]{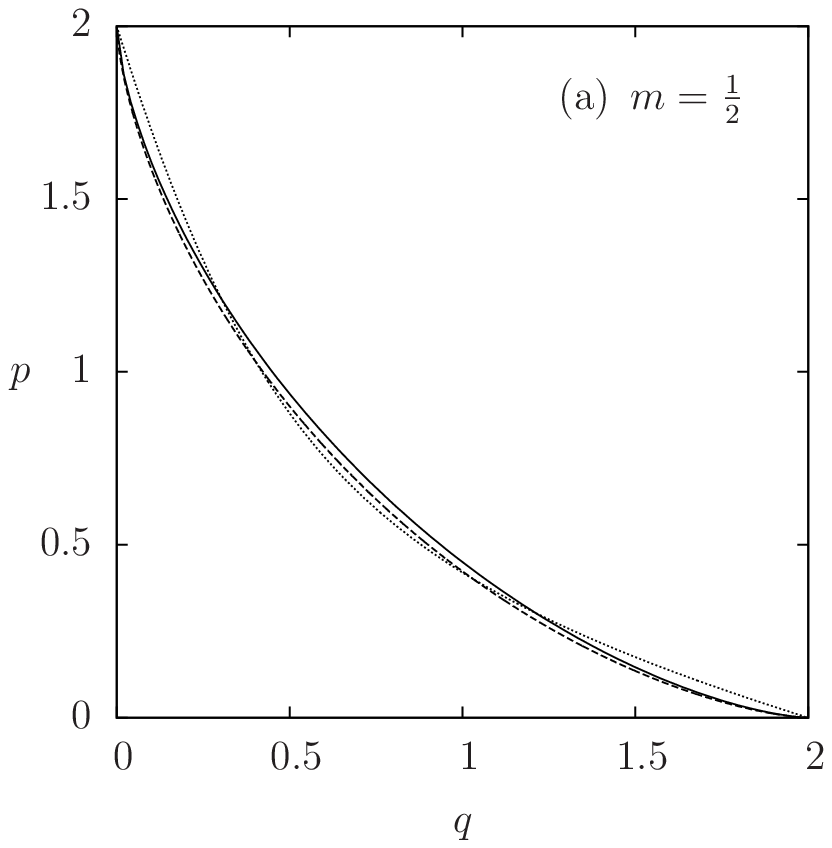}
\includegraphics[width=3in]{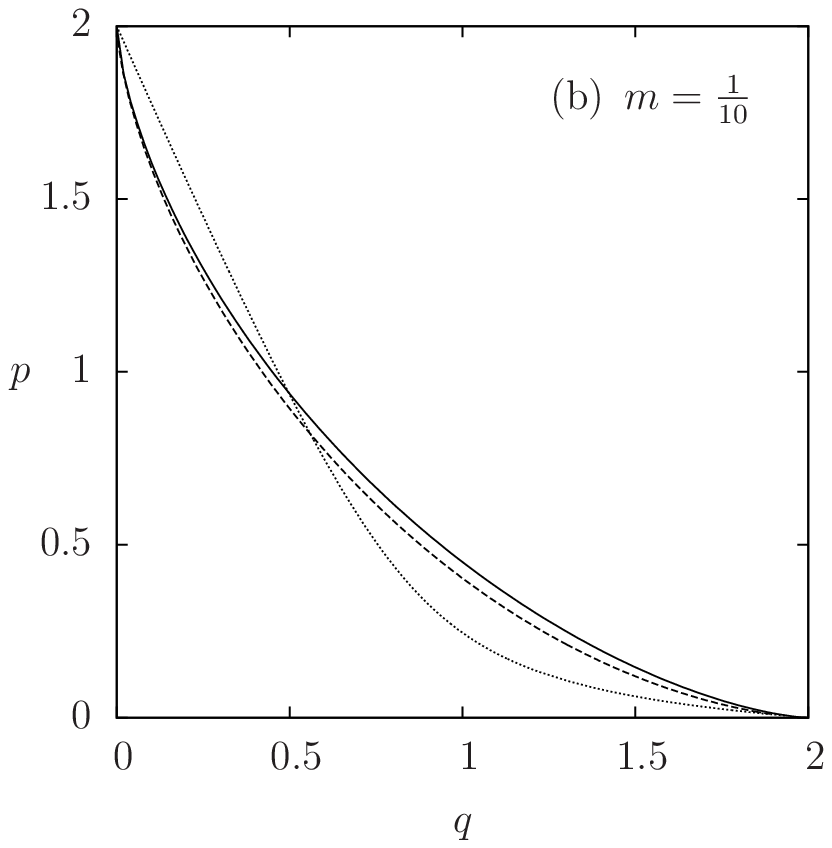}
\includegraphics[width=3in]{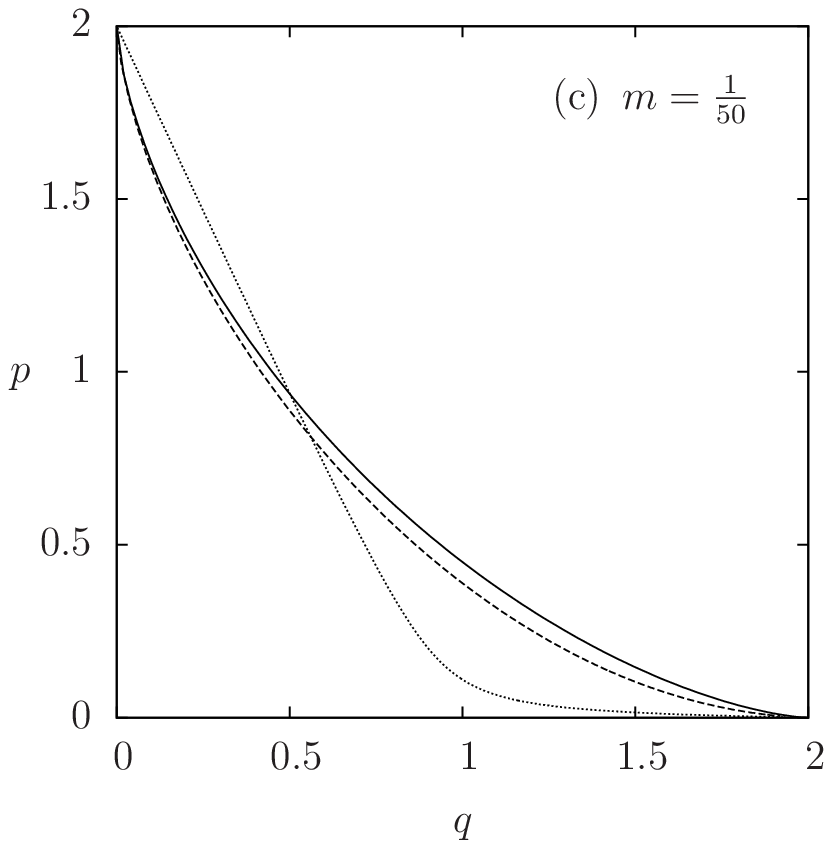}
\includegraphics[width=3in]{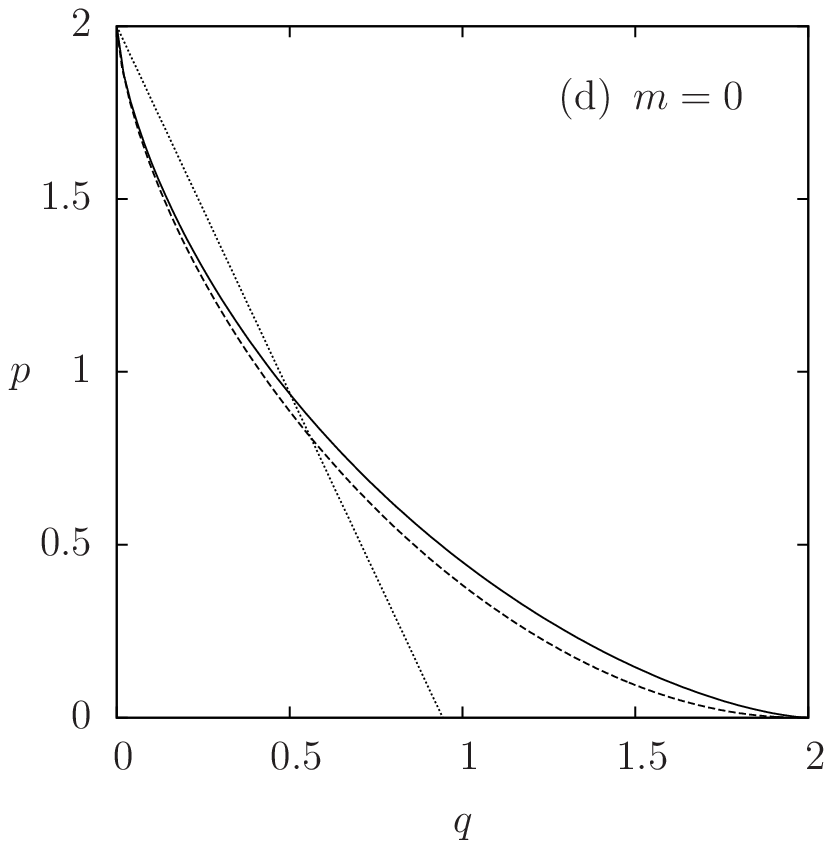}
\caption{The phase space $p$ versus $q$ of the balanced dihole system for the cases (a)~$m=\frac{1}{2}$, (b)~$m=\frac{1}{10}$, (c)~$m=\frac{1}{50}$, and (d)~$m=0$. In each case, three representative curves with fixed $l$ are plotted: the values of $l$ for the solid, dashed and dotted curves are $0$, $1$, and $\infty$ respectively.}
\label{fig_phase_space}
\end{center}
\end{figure}

It is perhaps even more interesting to see how the phase space, rather than the parameter space, looks like. Since the angular momentum of the solution is not an independent quantity, if we use the total mass as a common scale, the phase space is then characterised by the electric and magnetic charges. The dimensionless electric and magnetic charges of the solution are defined by and calculated to be
\ba
q&\equiv&\frac{|Q_{\rm e}|}{M_{\rm tot}}={\frac { 2\left( a+l+1 \right) \sqrt {a(a+1-m)}}{(a+b+2)\sqrt {(a+l+1-m)(a+b+l+1
)}}}\,,\cr
p&\equiv&\frac{|P_{\rm m}|}{M_{\rm tot}}={\frac { 2\left( b+l+1 \right) \sqrt {b(b+m)}}{(a+b+2)\sqrt {(b+l+m)(a+b+l+1
)}}}\,.
\ea
The balance condition then imposes a constraint on $p$ and $q$ for fixed $l$ and $m$, which is of course expected. Three representative $p$ versus $q$ curves for fixed $l$ are plotted in Fig.~\ref{fig_phase_space}(a), for the case $m=\frac{1}{2}$. The first is a solid curve with $l\rightarrow 0$, and is given by the equation
\be
\label{merging_curve}
p^{\frac{2}{3}}+q^{\frac{2}{3}}=2^{\frac{2}{3}}.
\ee
Curves with small fixed $l\ll1$ are well represented by this curve. Note that in the case $l\rightarrow 0$, we have $a,\,b\rightarrow \infty$, which means that the irreducible masses of the two black holes and the distance $l$ between them (in units of $z_{32}+z_{54}$) are much less than their electric and magnetic charges. So this limit can be understood as the merging {\it and\/} extremal limit of the present solution.\footnote{As we shall see in Sec.~6.5, the solely merging limit will give a rotating dyonic Kaluza--Klein black hole \cite{Rasheed:1995,Larsen:1999} with $J=PQ$. The merging and extremal limit of the present solution will then give the extremal limit of this Kaluza--Klein black hole, which can be identified with the intersection curve of the \textbf{S} and \textbf{W} surfaces in Fig.~2 of \cite{Rasheed:1995}.} The second, which is the dashed curve in Fig.~\ref{fig_phase_space}(a), has $l=1$. The third curve, corresponding to $l\rightarrow \infty$, represents the case for large $l\gg1$. It is drawn as a dotted curve in the same figure. Note that this curve intersects each of the previous two curves twice. The intersection points represent phases with two-fold non-uniqueness within the present case. Such intersecting behaviour generally occurs for a curve with a large enough $l$ and one with a small enough $l$.

We have also plotted in Figs.~\ref{fig_phase_space}(b), (c) and (d) representative $p$ versus $q$ curves for the cases $m=\frac{1}{10}$, $\frac{1}{50}$ and $0$, respectively. In each of these cases, it can be seen that the $l\rightarrow0$ solid curve is again given by (\ref{merging_curve}). However, the $l=1$ dashed curve and especially the $l\rightarrow\infty$ dotted curve begin deforming in response to the change in $m$, and both are no longer symmetrical under interchange of the $p$- and $q$-axes. In particular, when $m=0$, it can be shown that the $l\rightarrow\infty$ dotted curve actually becomes the piecewise linear curve given by
\be
\label{piecewise_curve_1}
p=\left\{\begin{array}{ll}
2-\hbox{$\frac{3}{\sqrt{2}}$}\,q\,,\qquad&\hbox{if\, $0\leq q\leq\frac{2\sqrt{2}}{3}$}\,;\\
0\,,\qquad &\hbox{if\, $\frac{2\sqrt{2}}{3}< q\leq2$}\,.
\end{array}\right.
\ee

We note that the cases $m=\frac{9}{10}$, $\frac{49}{50}$ and $1$ can be simply obtained from Figs.~\ref{fig_phase_space}(b), (c) and (d) respectively by interchanging the $p$- and $q$-axes. In particular, when $m=1$, the $l\rightarrow\infty$ dotted curve becomes the piecewise linear curve given by 
\be
\label{piecewise_curve_2}
q=\left\{\begin{array}{ll}
2-\hbox{$\frac{3}{\sqrt{2}}$}\,p\,,\qquad&\hbox{if\, $0\leq p\leq\frac{2\sqrt{2}}{3}$}\,;\\
0\,,\qquad &\hbox{if\, $\frac{2\sqrt{2}}{3}< p\leq2$}\,.
\end{array}\right.
\ee

\begin{figure}[t]
\begin{center}
\includegraphics[width=3in]{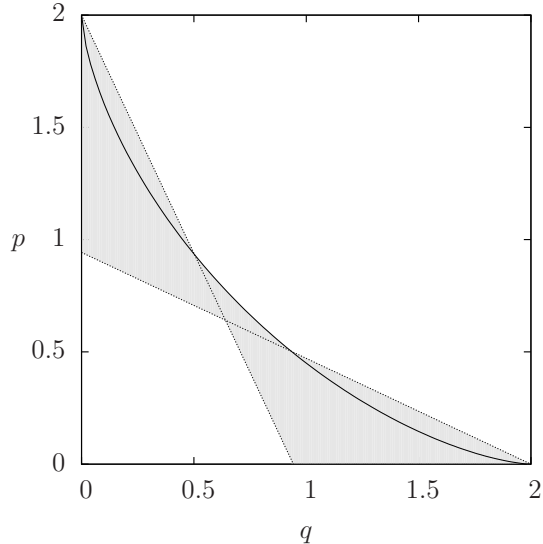}
\caption{Plot of the whole phase space \textbf{X}, with the solid curve (\ref{merging_curve}) and the two dotted curves (\ref{piecewise_curve_1}) and (\ref{piecewise_curve_2}) plotted for reference. The phase space \textbf{Y} of the Kaluza--Klein black hole of \cite{Rasheed:1995,Larsen:1999} with $J=PQ$ is the area bounded by the solid curve and the $p$- and $q$-axes. Note that this figure is plotted on the same scale as those in Fig.~\ref{fig_phase_space}.}
\label{fig_whole_phase_space}
\end{center}
\end{figure}

It is instructive to see what the whole $p$-$q$ phase space (denoted by \textbf{X}) looks like, for all possible values of $l$ and $m$. This is shown as the shaded area in Fig.~\ref{fig_whole_phase_space}, and is bounded by the solid curve (\ref{merging_curve}) and the two piecewise linear dotted curves (\ref{piecewise_curve_1}) and (\ref{piecewise_curve_2}). As expected, it is symmetrical under interchange of the $p$- and $q$-axes. It is also clear that there is a high degree of non-uniqueness of the phases in \textbf{X}. Besides the discrete non-uniqueness for fixed $m$ as observed above, there is a continuous non-uniqueness parameterised by $m$ itself. For example, consider points in the $m=\frac{1}{2}$ phase space (Fig.~5(a)) lying along the diagonal $p=q$. It can be checked that these points also lie in the phase space for every other $m\neq\frac{1}{2}$. Thus, these points in \textbf{X} have a continuous non-uniqueness parameterised by $m$ for every value in the interval $(0,1)$. Such continuous non-uniqueness will also extend to other points of \textbf{X}, although in general the range of $m$ will be smaller than that above.

It is worthwhile to compare \textbf{X} with the phase space (denoted by \textbf{Y}) of the rotating dyonic Kaluza--Klein black hole \cite{Rasheed:1995,Larsen:1999} with $J=PQ$. \textbf{Y} covers exactly the area bounded by the solid curve and the $p$- and $q$-axes. The solid curve in \textbf{Y} now represents the extremal limit of the Kaluza--Klein black hole, in complete agreement with the fact that this curve in \textbf{X} corresponds to $l\rightarrow 0$, representing the merging and extremal limit of the present solution. It is clear that \textbf{X} and \textbf{Y} have a non-empty intersection. Every point in this intersection describes a family of balanced diholes that shares the same phase as a Kaluza--Klein black hole. 

Another interesting point to note is that this non-uniqueness property exists even if the total scalar charge (in certain ranges) is specified. To see this, consider the intersection between the $m=\frac{1}{2}$ phase space, \textbf{Y} and the diagonal $p=q$, which is obviously non-empty. Points in this intersection have vanishing total scalar charge, and yet there is a two-fold non-uniqueness, with one phase in \textbf{X} and the other in \textbf{Y}. Hence, it is possible for two solutions in Kaluza--Klein theory to have the same total mass, angular momentum, scalar charge, and electric and magnetic charges, but yet they describe physically different configurations.

\newsection{Far-separation approximation}

When the two four-dimensional black holes are far-separated, we can define their individual masses, angular momenta and scalar charges, and treat the interaction between them in the Newtonian approximation. The individual quantities of a black hole are defined as the ones when the other black hole is set to be vanishing and pushed to infinity. The latter limit is taken as $z_{43}\rightarrow \infty$. Note that in this limit, $z_{43}$ can be identified as the proper distance $r$ between the two black holes. Direct calculations then yield
\ba
M_{\rm e}=\frac{z_{31}+z_{32}}{4}\,,\qquad J_{\rm e}&=&0\,,\qquad \Sigma_{\rm e}=\frac{\sqrt{3}z_{21}}{4}\,,\cr
M_{\rm m}=\frac{z_{54}+z_{64}}{4}\,,\qquad J_{\rm m}&=&0\,,\qquad \Sigma_{\rm m}=-\frac{\sqrt{3}z_{65}}{4}\,.
\ea
In particular, we see that the two black holes carry scalar charges with opposite signs. In this limit, the electric and magnetic charges also simplify: $Q_{\rm e}=\frac{\sqrt{z_{21}z_{31}}}{2}$, $P_{\rm m}=-\frac{\sqrt{z_{64}z_{65}}}{2}$. Note that in this approximation both $J_{\rm e}$ and $J_{\rm m}$ vanish, and they do not sum up to $J_{\rm tot}$. This can be attributed to the fact that the angular momentum of the system in this limit is entirely stored in the electromagnetic field.

The total force between the two black holes can be written as
\be
\label{force_formula1}
F_{\rm int}=-\frac{M_{\rm e}M_{\rm m}}{r^2}-\frac{\Sigma_{\rm e}\Sigma_{\rm m}}{r^2}\,.
\ee
Since one of the black holes is purely electric, and the other purely magnetic, there is no direct electromagnetic interaction in the above force formula. Note that the dilatonic force is repulsive if the two scalar charges have opposite signs, as in the present solution. For the present solution, the total force can be written as
\be
\label{force_formula2}
F_{\rm int}=\frac{3z_{21}z_{65}-(z_{21}+2z_{32})(z_{65}+2z_{54})}{16r^2}\,.
\ee

In this limit it is easy to show that, to leading order in $\frac{1}{r}$, the balance condition (\ref{balance_condition}) reduces to
\be
\label{balance_condition2}
F_{\rm int}=0\,,
\ee
which is of course the condition for equilibrium in the Newtonian approximation. If this condition is satisfied, the repulsive dilatonic force cancels gravity exactly at this order. This means that the system is actually in neutral equilibrium to leading order, and radial instability is at most a next-to-leading order effect.

It is instructive to examine a few cases where the condition (\ref{balance_condition2}) cannot be satisfied. Let us first consider the extremal limit $z_{32}=0$ and $z_{54}=0$. In this case, the total force reduces to
\be
F_{\rm int}=\frac{z_{21}z_{65}}{8r^2}> 0\,,
\ee
which indicates that the system cannot be balanced in this configuration, and the repulsive dilatonic force overwhelms gravity. So a conical singularity is necessarily present. We point out, however, that a balanced system corresponding to an extremal electric-magnetic dihole in string theory is possible \cite{Kallosh:1992ii}. This is because in this case the dilaton coupling constant is smaller than in Kaluza--Klein theory, and the repulsive dilatonic force is thus weaker, so that it is possible to cancel gravity exactly.

It is also clear from the above force formula that the system cannot be balanced if either the electric or the magnetic charge is turned off, i.e., if either $z_{21}=0$ or $z_{65}=0$. Although the electric and magnetic charges themselves do not appear explicitly in the force formula, they do set the upper (absolute) bounds on the scalar charges that the black holes can carry:
\be
|\Sigma_{\rm e}|\leq \frac{\sqrt{3}}{2}\,|Q_{\rm e}|\,, \qquad |\Sigma_{\rm m}|\leq \frac{\sqrt{3}}{2}\,|P_{\rm m}|\,.
\ee
So if the electric or magnetic charges are too small, one or both of the scalar charges are necessarily small, and the repulsive dilatonic force cannot balance gravity.

\newsection{Various limits}

\newsubsection{Emparan--Reall black ring on Taub-NUT}
\label{subsection_ER_TN}

In this limit, the black hole of the black-Saturn system vanishes, leaving an Emparan--Reall black ring on Taub-NUT discovered by the present authors \cite{Chen:2012zb}. It is obtained straightforwardly from (\ref{metric_bs_tn}) by setting
\be
z_4\rightarrow z_5\,.
\ee
To map the resulting solution to the exact form presented in \cite{Chen:2012zb}, we need to perform the following parameter redefinitions and coordinate transformations:
\ba
z_1=-\frac{2b-c(1+b)}{1-b}\,\varkappa^2,\qquad z_2=-c\varkappa^2,\qquad z_3=c\varkappa^2,\qquad z_{4,5}=\varkappa^2,\qquad z_6=a\varkappa^2,\cr
\rho = \frac{2\varkappa^2 \sqrt{(1-x^2)(y^2-1)(1+cx)(1+cy)}}{(x-y)^2}\,,\qquad
 z   =\frac{\varkappa^2 (1-xy) (2+cx+cy)}{(x-y)^2}\,,\hskip.15in
\ea
followed by the replacement $\psi\rightarrow \psi-2n\phi$, where $n$ is the NUT charge in this limit.

\newsubsection{Static black hole on Taub-NUT}

We recover a static black Saturn on Taub-NUT from (\ref{metric_bs_tn}) by setting
\be
\label{zero_electric_charge}
z_1\rightarrow z_2\,.
\ee
In Kaluza--Klein theory, this solution describes a superposition of two static black holes, with one of them neutral and the other magnetically charged. As we have mentioned, when the electric charge (and so the rotation of the black ring from the five-dimensional perspective) of the left black hole is sufficiently small, the balance condition cannot be satisfied. This is of course the case for the current situation: Since there is no repulsive force resulting from the dilaton-dilaton interaction, a conical (strut) singularity must be present in between the two black holes to balance gravity. It is direct to show that $\kappa_{\rm E}>1$ in this case.

If we further make the black ring of the black Saturn system vanish, by setting
\be
z_2\rightarrow z_3\,,
\ee
we obtain a static black hole on Taub-NUT. In this limit case, all the expressions $z_1,z_2,z_3$ become irrelevant, and the conical singularity disappears. We then define
\ba
z_4=-\frac{r_2-r_1}{2}\,,\qquad z_5=\frac{r_2-r_1}{2}\,,\qquad z_6=\frac{r_1+r_2}{2}\,,\hskip0.6cm\cr
\rho=\sqrt{(r-r_1)(r-r_2)}\,\sin\theta\,,\qquad z=\Big(r-\frac{r_1+r_2}{2}\Big)\cos\theta\,,
\ea
followed by the replacement $\psi\rightarrow \psi-\sqrt{r_1r_2}\,\phi$. The metric finally becomes
\ba
\dif s^2&=&\left(1-\frac{r_1}{r}\right)\left[\dif \psi+\sqrt{r_1r_2}\cos\theta\,\dif \phi\right]^2-\frac{1-\frac{r_2}{r}}{1-\frac{r_1}{r}}\,\dif t^2+\left(1-\frac{r_2}{r}\right)^{-1}\dif r^2\cr
&&+\left(1-\frac{r_1}{r}\right)r^2(\dif \theta^2+\sin^2\theta\,\dif \phi^2)\,.
\ea
One immediately recognises this to be the metric of the static magnetic Kaluza--Klein black hole lifted to five dimensions, which is equivalent to that of the static black hole on Taub-NUT.

\newsubsection{Zero NUT-charge limit}

In this limit, the NUT charge of the solution becomes zero. It is taken by setting
\be
\label{zero_NUT_limit}
z_5\rightarrow z_6\,.
\ee
In view of the identifications made in (\ref{identifications_coor}), the size of the compact dimension generated by $\frac{\partial}{\partial\psi}$ turns out to vanish at infinity in this limit.\footnote{Essentially the same feature is also present in the zero-NUT charge limit of Taub-NUT space, as previously discussed in \cite{Chen:2010zu}.} We remark, however, that the second identification in (\ref{identifications_coor}), being identical to the first, is no longer necessary to ensure regularity of the resulting space-time; in fact, we can now make an arbitrary identification on the orbits generated by $\frac{\partial}{\partial\psi}$. If we identify it with a finite period, the resulting metric describes a configuration of two black rings/strings in five dimensions, with one of them rotating in the compact dimension.

In Kaluza--Klein theory, this limit describes a superposition of two static black holes, with one of them electrically charged and the other neutral. As with the static black-Saturn limit interpreted in Kaluza--Klein theory as described in the previous subsection, it is direct to show that $\kappa_{\rm E}>1$ in this case. Since there is no repulsive dilatonic force emerging, a conical (strut) singularity must be present to balance gravity.
Note that a special case of this solution arises when we further take the limit (\ref{zero_electric_charge}), in which we recover the Israel--Khan solution \cite{Israel} describing a superposition of two Schwarzschild black holes.

\newsubsection{Elvang--Figueras black Saturn}

In the infinite NUT-charge limit, the solution (\ref{metric_bs_tn}) becomes asymptotically flat, and we recover the black Saturn of Elvang and Figueras \cite{Elvang:2007rd}. This is obtained by defining
\be
t=\sqrt{4n}\,\tilde{t}\,,\qquad \psi=-4n\tilde{\psi}\,,\qquad \phi=\tilde{\psi}+\tilde{\phi}\,,
\ee
taking out an overall factor $4n$, and then setting
\be
z_6\rightarrow \infty\,.
\ee
The limiting solution is then equivalent to the solution presented in \cite{Elvang:2007rd} with $c_2=0$, although it is simpler in form. To show their equivalence we need to identify $z_1,...,z_6$ with the positions of the turning points from left to right in \cite{Elvang:2007rd}, use the identity $R_{ij}\mu_{ij}=2z_{ij}\mu_i\mu_j$, and identify the coordinates $(\tilde{t},\tilde{\psi},\tilde{\phi})$ with $(t,\psi,\phi)$ used there.

\newsubsection{Merging limit}

The distance between the black hole and black ring in the black-Saturn system is largely characterised by the length $z_{43}$. When this quantity is very small, it can be seen from (\ref{kappa_E}) that the conical singularity has an excess angle that becomes very large. Physically, this means that gravity dominates the system when the black ring is very near the black hole, so that the other interactions cannot balance the system and a conical (strut) singularity must be present. In the limit
\be
z_4\rightarrow z_3\,,
\ee
the length of the conical singularity shrinks down to zero, even as the excess angle diverges, and the black ring and black hole actually merge into a single bigger object.

The resulting solution can be shown to be a singly rotating black hole on Taub-NUT, with no rotation along the direction $\ell_{\rm I}$. In Kaluza--Klein theory, it describes a rotating dyonic black hole \cite{Rasheed:1995,Larsen:1999} with $J=PQ$. To map the limiting solution to the latter form, we first need to reparameterise the positions of relevant turning points by
\be
\quad z_1=-\frac{pq^2-2m^2(p-q)}{pq+4m^2}\,,\qquad z_{2,5}=\mp\frac{2m^2(p+q)}{pq+4m^2}\,,\qquad z_6=\frac{p^2q+2m^2(p-q)}{pq+4m^2}\,,
\ee
and then define the coordinates $(r,\theta)$ by
\be
\rho=\left[r^2-2mr+\frac{m^2(p^2-4m^2)(q^2-4m^2)}{(pq+4m^2)^2}\right]^{\frac{1}{2}}\sin\theta\,,\qquad z=(r-m)\cos\theta\,.
\ee
The final solution, after the replacement $\psi\rightarrow \psi-2n\phi$, where $n$ is the NUT charge in this limit, can be shown to be exactly the same as the rotating dyonic Kaluza--Klein black hole with $J=PQ$, in the five-dimensional form given by Larsen \cite{Larsen:1999}.\footnote{C.f.~Footnote 1 of \cite{Chen:2012zb} for a remark on the conventions of Larsen vis-\`a-vis ours.}

\newsubsection{Extremal limit}

With the present parameterisation, the extremal limit of the solution (\ref{metric_bs_tn}) is fairly easy to take. It is obtained by directly setting
\be
\label{extremal_limit}
z_2\rightarrow z_3\,,\qquad z_4\rightarrow z_5\,.
\ee
In fact, this limit can be taken step by step: Firstly, we can take the second limit in (\ref{extremal_limit}) to make the black hole vanish, which of course results in the Emparan--Reall black ring on Taub-NUT as discussed in Sec.~6.1. Then we can take the first limit to make the black ring extremal. This extremal limit is thus seen to be equivalent to the extremal limit of the Emparan--Reall black ring on Taub-NUT discussed in \cite{Chen:2012zb}, which is in turn equivalent to the extremal limit of the solution found in \cite{Camps:2008hb}. We refer the reader to these two papers for more details of the extremal limit. We remark that in this limit, both black holes in Kaluza--Klein theory actually become singular with vanishing area, and a conical singularity is necessarily present to balance the system.

\newsection{Discussion}

To summarise the main results of this paper, we have used the ISM to construct a black Saturn on Taub-NUT in five-dimensional vacuum gravity. This system consists of an $S^1$-rotating black ring around a static black hole, and can be balanced for appropriately chosen parameters. When this solution is reduced to four dimensions, it describes a balanced system consisting of an electrically charged black hole and a magnetically charged black hole. Such an electric-magnetic dihole system is, to the best of our knowledge, the first known way of achieving equilibrium in an asymptotically flat, non-extremal and non-supersymmetric multi-black-hole configuration in four dimensions. It also provides a clear example of black-hole non-uniqueness in four dimensions, at least in Kaluza--Klein theory: for certain ranges of the asymptotic conserved charges, there exists a continuous infinity of physically different configurations sharing the same conserved charges.

Although we have analysed the key properties of the electric-magnetic dihole solution, there remain a few open questions regarding its properties that could be the focus of future investigations. For example, we have seen in Sec.~4 that the dihole system can be balanced at any distance (for appropriate electric and magnetic charges), and not just for the case of large separation subsequently analysed in Sec.~5. For smaller separations, other effects besides the gravitational and dilatonic forces are expected to come into play, including possible effects arising from the non-zero angular momentum of the space-time. To come to a complete understanding of why balance is possible at any distance, as well as to gain insight into the overall stability of the system, it would be necessary to use different methods to isolate and study the various effects involved.

Even if the system is balanced, the two black holes, being a finite distance apart, should still feel the effects of each other. This might be manifest, for example, as deformations of the event horizons away from spherical symmetry. The shape of the ergo-surfaces are also expected to vary with the separation of the black holes, and there might even be a merging of the ergoregions \cite{Elvang:2008qi} if the black holes are sufficiently close together. It would be interesting to study these issues, say along the lines of what has been done for the double-Kerr solution \cite{Herdeiro:2008kq,Costa:2009wj}.

It is our hope that this solution will pave the way for the discovery of other examples of balanced electric-magnetic dihole systems. As a first attempt, it should be possible to add rotation to the individual black holes of our solution while still maintaining balance. The individual black-hole angular momenta will then contribute to the total angular momentum of the space-time. In the five-dimensional picture, this would correspond to making the black ring rotate in both possible directions, and the black hole rotate in an appropriate direction. It should be straightforward to construct this solution using the ISM.

It should also be possible to generalise the solution of this paper to include an arbitrary number of electrically charged black holes in balanced superposition with the magnetically charged black hole along a common axis. In the five-dimensional picture, this would correspond to a number of black rings, rotating in the same plane around a central static black hole. Again, such a solution can in principle be constructed using the ISM, although it is likely to be very complicated.

A more intriguing question is whether balanced electric-magnetic diholes can exist in other theories, such as Einstein--Maxwell-dilaton theory with other values of dilaton coupling. For pure Einstein--Maxwell theory, there will not be a dilatonic force available to balance the gravitational force at large separations. However, it is conceivable that some other shorter-ranged effect might still be able to balance gravity at small separations, enabling a balanced dihole to exist in this theory. 
Another theory to consider that is somewhat more similar to Kaluza--Klein theory, is Einstein--Maxwell-dilaton theory with a string coupling, such as that considered in \cite{Kallosh:1992ii}. It might be worthwhile to see if a balanced, non-extremal dihole can exist in this theory.

Returning to four-dimensional Kaluza--Klein theory, we note that there is already an interesting string-theory interpretation of the solution of this paper. Recall that extremal electric and magnetic black holes in this theory can be interpreted as D0 and D6 branes respectively, when this theory is embedded in 10-dimensional Type-IIA string theory. Our non-extremal dihole system then corresponds to a non-supersymmetric system of thermally excited D0 and D6 branes, separated by a finite distance. It would be interesting to carry out a study of this D-brane system.

\bigbreak\bigskip\bigskip\centerline{{\bf Acknowledgements}}
\nobreak\noindent We are grateful to Roberto Emparan and Carlos Herdeiro for their comments and suggestions on the manuscript. We are also grateful to the referee for some useful suggestions on the first version of this paper. YC wishes to thank Joan Camps for useful discussions. ET wishes to acknowledge the kind hospitality of the Centre for Gravitational Physics, where this work was carried out. This work was supported by the Academic Research Fund (WBS No.: R-144-000-277-112) from the National University of Singapore.

\appendix

\newsection{Smoothness of the space-time}

As mentioned, the space-time described by the metric (\ref{metric_bs_tn}) is well-behaved on and outside the event horizons if the balance condition (\ref{balance_condition}) is imposed. Smoothness of the metric can be proved along the lines of \cite{Chrusciel:2010ix}, and in this appendix, we will present the main points of this proof. To simplify matters, we focus on the case when all the parameters $z_i$ (for $i=1,...,6$) are distinct, satisfying the range $z_i>z_j$ if $i>j$. Similar analyses can be carried out in the various limits when two or more of these parameters coincide. We also assume for the moment that we are working from the five-dimensional perspective.

Firstly, we prove that away from the axes and outside the event horizons, the space-time is free of CTCs. This requires that the two-dimensional metric $g_{ab}$ for the vector space spanned by the basis $\frac{\partial}{\partial x^a}$ for $x^{a}=(\psi,\phi)$ is positive definite. To see this, we write the metric in a form more tractable for analysis as follows:
\be
\label{metric_psiphi}
g_{ab}\,\dif x^a \dif x^b=\frac{\mu_1\mu_2\mu_5\mu_6 K}{H}\left(\dif \psi+\omega_2\,\dif \phi\right)^2+\frac{\mu_3 \rho^2L}{\mu_2\mu_4\mu_6 K}\,\dif \phi^2,
\ee
where the functions $\omega_2$, $H$, $K$ are those in (\ref{functions_A}) and (\ref{functions_HF}), and $L$ is given by
\be
L=-C_1^2C_2^2\rho^6M_1-C_1^2\mu_6^2M_2-C_2^2\rho^4M_3+\mu_6^2M_4\,.
\ee
Since $\mu_i$ is always positive in the region under consideration, the task is now to prove that $H$, $K$ and $L$ are positive. Recalling the identity $R_{ij}\mu_{ij}=2z_{ij}\mu_i\mu_j$, we first write the constants $C_1^2$ and $C_2^2$ in the particular form
\be
C_1^2=\frac{\mu_1^4\mu_6\mu_3\mu_2R_{14}\mu_{41}\mu_5^2}{R_{16}\mu_{61}R_{13}\mu_{31}R_{12}\mu_{21}\mu_4R_{15}^2\mu_{51}^2}\,,\qquad C_2^2=\frac{\mu_6^4\mu_5\mu_4\mu_1R_{36}\mu_{63}\mu_2^2}{R_{56}\mu_{65}R_{46}\mu_{64}R_{16}\mu_{61}\mu_3R_{26}^2\mu_{62}^2}\,,
\ee
and substitute them into the functions $H$, $K$ and $L$. Now these three functions contain expressions in terms of only $\mu_i$ and $\rho$ (recalling that $\mu_{ij}=\mu_i-\mu_j$). Next, we express the variables $\mu_i$ in terms of $\mu_1$ and $\mu_{i+1,i}$ (for $i=1,...,5$), all of which are obviously positive quantities. It is then direct to show that the numerators and denominators of the functions $H$, $K$ and $L$ are polynomials in terms of the new variable-set $(\rho,\mu_1,\mu_{i+1,i})$, and all the coefficients of their individual terms are positive. This finishes the proof. This result, in particular, implies that $g_{\psi\psi}$ is positive, which will be needed below.

Secondly, it is now straightforward to show that the metric (\ref{metric_bs_tn}) is smooth and has the correct signature away from the axes and event horizons. We begin by noting that the metric components $g_{\mu\nu}$ ($\mu$ and $\nu$ run over all coordinate indices) are functions of $\rho$ and $\mu_i$ only, all of which are smooth functions in terms of the coordinates $(\rho,z)$ (for $\rho>0$). Moreover, the denominators of the metric components contain only (a) factors of the obviously positive quantities $R_{ij}$, $\mu_i$ and $\mu_{ij}$ with $i>j$, and/or (b) a factor of the longer quantity $H$, which has also been shown to be positive. It follows that the metric is smooth in the region under consideration. To see that the metric has the correct signature, we first observe that the two-dimensional diagonal part of the metric $g_{\rho\rho}\,\dif\rho^2+g_{zz}\,\dif z^2$ is positive definite. Furthermore, the determinant of the metric is a non-zero smooth function, which is an easy task to show since direct calculations yield $\det g_{\mu\nu}=-\rho^2 R^2H^2$. Together with the asymptotic structure (\ref{asymptotic_structure}), it is then clear that the solution has Lorentzian signature away from the axes and event horizons.

Now we proceed to prove smoothness of the space-time on each axis and each horizon, excluding the points where they meet. This is partially settled in the rod-structure formalism \cite{Harmark:2004rm,Hollands:2007aj,Chen:2010zu}. In this formalism, these regions are represented by the interior points of the rods; in the present case, they are the {\it open\/} intervals $(-\infty,z_2)$, $(z_2,z_3)$, $(z_3,z_4)$, $(z_4,z_5)$ and $(z_5,\infty)$ along the $z$-axis. The metric behaviour in the interior of a rod has been studied in say \cite{Harmark:2004rm}. Recall that we have joined up Rods 1 and 2, as well as Rods 6 and 7 in our solution, so that the points $z_{1},z_{6}$ are not rod end-points anymore---they are now interior points of Rods I and V respectively. These points are called phantom points in the literature, and can be viewed as turning points that have been eliminated from the rod structure, although they do continue to serve as independent parameters of the solution.

In each open interval, $g_{\rho\rho}=g_{zz}$ can be shown to be finite, positive and smooth. We illustrate this with a representative interval, say for $z_2<z<z_3$. In this case, we express $g_{\rho\rho}$ in terms of $z-z_2$, $z_3-z$, and $z_{i+1,i}\equiv z_{i+1}-z_i$ for $i=1,3,4,5$. By inspection it is clear that each term of the numerator and denominator is positive, and hence $g_{\rho\rho}$ is positive. It is then direct to show that all the other metric components of (\ref{metric_bs_tn}) are finite and smooth along the $z$-axis. In particular, $g_{\psi\psi}$ can be shown to be finite and positive. In each open interval, the determinant of the metric becomes zero, as can be seen from the definition of the rods in the rod structure. This is expected, since we can show that the space-time closes off smoothly in the interior of an axis-rod and can be smoothly extended through the interior of a horizon-rod (see, e.g., \cite{Chrusciel:2010ix}). CTCs are also absent in these regions, due to the continuity of the two-dimensional metric $g_{ab}$ and thus its eigenvalues.

Now we show that the conditions (\ref{C12}) are also consequences of the smoothness requirement. The value of $C_{1}$ in (\ref{C12}) has been chosen as such to ensure that the metric components do not blow up at $z_{1}$; if instead $C_1$ were not chosen as in (\ref{C12}) or its negative, the metric component $g_{\psi\psi}$ will diverge and a singularity will appear at $z_1$. A general analysis of the removable nature of a phantom-point singularity such as $z_1$ can be found in \cite{Chrusciel:2010nz}. On the other hand, $g_{\psi\psi}$ will be zero at $z=z_6$ unless $C_2$ takes the value in (\ref{C12}) or its negative. Hence, the value of $C_2$ has been chosen as such to ensure that $g_{\psi\psi}\ne 0$ at $z_6$, thus avoiding a naked singularity there when dimensional reduction is performed along the direction $\frac{\partial}{\partial \psi}$ in Kaluza--Klein theory. We believe that the removable nature of a phantom point such as $z_6$ in Kaluza--Klein theory can also be proved in a way similar to \cite{Chrusciel:2010nz}.

Lastly, we need to examine the space-time geometry around the four genuine turning points $z_2,z_3,z_4,z_5$, where the above-mentioned five rods meet. These points are where the two event horizons meet the axes. When such a point say $z_2$ is approached along the $z$-axis, the metric components $g_{\rho\rho}=g_{zz}$ become divergent, with $g_{\rho\rho}|z-z_2|$ remaining finite and continuous at $z_2$. All the other metric components are finite and smooth at this point along the $z$-axis. In particular, $g_{\psi\psi}$ is finite and positive. Following \cite{Chrusciel:2010ix}, one can prove that the divergence of the metric components $g_{\rho\rho}=g_{zz}$ just indicates that the Weyl--Papapetrou coordinates break down at these points, and by finding appropriate coordinates, the metric can be shown to be smooth around these points.

We end this appendix by remarking that the solution is also smooth on and outside the event horizons when interpreted in Kaluza--Klein theory after dimensional reduction along the direction $\frac{\partial}{\partial \psi}$. This can be seen from the above-mentioned fact that $g_{\psi\psi}$ is always positive and finite in the region under consideration. Absence of CTCs in the five-dimensional space-time, resulting from the positive semi-definiteness of the metric (\ref{metric_psiphi}), then ensures the absence of CTCs in the reduced four-dimensional space-time (\ref{metric_4D}).

\bigskip\bigskip

{\renewcommand{\Large}{\normalsize}
}

\end{document}